\documentclass[aps,pal,14pt,twocolumn,superscriptaddress,groupedaddress]{revtex4}  
\usepackage{amssymb}
\usepackage{amsmath}
\usepackage{amscd}
\usepackage{graphicx}
 \begin{document}
                                                                                
\title{Block entropy for Kitaev-type spin chains in a transverse field}
\author{ V. Subrahmanyam}
\email{vmani@iitk.ac.in}
\affiliation{ Department of Physics, Indian Institute of Technology, Kanpur-208016, India}
\date{\today} 
\begin{abstract}
Block entanglement entropy in the ground state of  a quantum  spin chain is investigated. The spins have  Kitaev-type nearest-neighbor interaction, of strength  $J_x$ or $J_y$, through either
x or y components of the spins  on  alternating bonds, along with a transverse magnetic field $h$.  An exact solution is obtained through
Jordan-Wigner fermionization, and it exhibits a macroscopically degenerate ground state for $h=0$, and  a non-degenerate ground state for $h\ne 0$, for all interaction strengths. For a chain of N spins, we study the block entropy of a partition of L contiguous spins. The block entanglement entropy  needs the
eigenvalues of the $2^{\rm L}$-dimensional reduced density matrix. We employ an efficient method that reduces this problem to evaluating eigenvalues
of a L-dimensional matrix, which enables us to calculate easily the block entanglement for large-N chains numerically. The entanglement entropy grows as
log L, at the degeneracy point $h=0$, and only for $J_x=J_y$. For $h\ne 0$, the entropy becomes independent
of the size, thus obeying the area law. For $J_x\ne J_y$, the block entropy shows a non-monotonic behavior for $\rm L<N/2$.

\hfill {PACS numbers:  03.67.Mn, 75.10.Pq,03.67.-a, 74.40.Kb}
\end{abstract}
\maketitle
 \noindent
\section{Introduction}
 
Entanglement in a bipartite quantum state,  which is a signature of the quantum correlations between the two parts, has been explored  from a wide range of viewpoints over last
few years.  Apart from the main thrust area of quantum computation and information processing, where quantum entanglement is recognized as a resource\cite{Nielsen},  there is
a growing interest in exploring entanglement properties of physical models exhibiting quantum critical-point phenomena\cite{plenio}, and conformal field theories\cite{srednicki}.  The von Neumann entropy of a subsystem, along with  its various monotones, is a quantitative measure of the entanglement between the subsystems\cite{vidal}. The entanglement entropy is not extensive as a function of the subsystem size, unlike physical observables.  
For systems with local interactions,  the correlations are expected to be short ranged (implying a energy-mass gap) in general,  and long ranged in the vicinity of a critical point (where
the correlation length becomes very large, or the energy gap vanishingly small). This leads to  the entanglement  entropy  obeying an area law in general, and near a critical point the
entropy exhibits a logarithmic dependence on the size. 

The study of entanglement in spin systems is dominated by one-dimensional models, with many different quantum spin models that can be exactly solved for the ground state and the excitation spectrum,  viz. XYZ chains using Bethe Ansatz, and XY chains with transverse magnetic field using Jordan-Wigner transformation\cite{vidal,plenio}.  In the context of fault-tolerant quantum
computation, Kitaev studied a two-dimensional honeycomb lattice spin model,  which
is solvable by the Jordan-Wigner fermion technique,   a rare example of exactly-solvable  two-dimensional system\cite{kitaev}.  This model shows remarkable exotic features, such as excitations obeying non-abelian statistics,  topological order and entanglement\cite{kitaev-e}. 
There have been generalizations to other lattices, and  to higher dimensions and higher spins\cite{general}. 
A one-dimensional variant of Kitaev model with no magnetic field has been explored for the phase diagram and the non-abelian anyon excitations analytically\cite{shanks}, for the quenched dynamics\cite{uma}. There are attempts to realize these spin models in systems of cold atoms\cite{cold} and superconducting quantum circuits\cite{qc}. 

Here, we will study the entanglement 
entropy of a simpler one-dimensional spin chain with Kitaev-type interactions  in the presence of
a transverse magnetic field. The one-dimensional model substantially simplifies the task of finding the ground state and further in obtaining the eigenvalues of the
reduced density matrix, as we will see below, but still retaining many of the exotic features of the Kitaev honeycomb model. We will employ a method of
operator Schmidt decomposition to compute the entanglement entropy. This method, which can be used in the context of other one-dimensional spin models, enables us to 
numerically compute the block entropy for spin chains with large number of lattice sites.
We describe the spin chain model and the exact solution from Jordan-Wigner transformation in the next section.  The ground state is written as a product state of $N/4$
modes, each mode consisting of four single-particle momentum states $q-\pi,-q,q,\pi-q$. In contrast, for the XY spin chain there are $N/2$ modes, each mode consisting of momentum $q$ and $-q$. In Section III, we show the Schmidt decomposition of the ground state. We express the ground state using an exponentiated operator,  and use
the method of operator Schmidt decomposition. This reduce the task of find the entanglement entropy to numerical diagonalization of a $N$ dimensional matrix. In the final
section we discuss the results from numerical computation of the block entanglement entropy in the Kitaev chain.  

 
\section{The model and the exact solution}
We consider a finite one-dimensional closed spin chain with  Kitaev-type interactions. Let
$\vec \sigma_i$ matrix vector represent  the spin-1/2 operator at a given site. The nearest-neighbor
spins interact  through either  x or y components of the spins on alternating pairs.  The  Hamiltonian for a closed chain of N (even) spins is written as,
\begin{equation}
H= \sum_{i =1,3,..}^{N-1} J_x \sigma_i^x \sigma_{i+1}^x   + \sum_{i=2,4..}^N J_y  \sigma_i^y \sigma_{i+1}^y +
h \sum_{i=1}^N \sigma_i^z,
\end{equation}
where $J_x$ and $J_y$ are interaction strengths on alternating bonds which favor x-x or y-y
polarization locally, and $h$ is the strength of the transverse magnetic field which favors z polarization
globally. Thus, this model cannot be mapped on to XY model with homogeneous couplings.
As we will see below,
this model has a macroscopic degeneracy in the ground state for all values of the interaction
strengths, for $h=0$. The degeneracy is completely lifted for a nonzero transverse magnetic field.
The ground state degeneracy can be lifted  by a staggered transverse field also. This implies
a large correlation length, and long-ranged spin correlations for $h=0$, and short-ranged correlations
for a nonzero field.   We will show later, that this does not translate directly into a logarithmically
increasing block entropy for $h=0$ and all interaction strengths.

Working with $\sigma_i^z$ basis for every spin, in a  basis state the possible values for the total z-component of the spin are, $\sigma^z = 0,1,2,3,..N$.
As the two interaction terms  in the above Hamiltonian are bilinear in x and y components of the spin operators, they mix only basis states within even (or odd) sector only.
But still, this global symmetry cannot account for a macroscopic degeneracy in the ground state for $h=0$, for all interaction strengths. We examine the
exact solution through Jordan-Wigner solution below. A more general spin chain has been investigated\cite{shanks}, in terms of unpaired Majorana modes and their interaction with the
background kink flux fields.  We will use the fermion language, as the
fermion operators can be more easily manipulated for calculating the block entropies.

The above Hamiltonian can be diagonalized,  by following the usual steps of exact diagonalization of spin chains\cite{Lieb}. All states can be
written as direct products of different fermion modes in momentum space. In XY model there are $ N/2$ uncoupled modes labeled by $q>0$, a  given mode mixing $\pm q$ momentum values. We will see below that we get here $N/4$ uncoupled modes,  with $0<q<\pi/2$,  mixing $q-\pi, -q, q, \pi-q$ momentum values.
We first go over to fermion variables from the spin operators that appear in the Hamiltonian using the Jordan-Wigner transformation\cite{Lieb}, given by
\begin{equation}
\sigma_l^z =2n_l-1,  ~\sigma_l^+ = e^{i\pi \sum_{j=1}^{l-1}n_j}c_l^{\dag},
\end{equation}
where $c^{\dag}_l$ creates a fermion at site $l$, and the number operator $n_l$ counts the fermions at the site. The $\sigma_i^z=\pm 1$ basis states at a site are mapped to occupied and unoccupied state of the fermion, $n_i=1,0$, respectively. Basis states with even $N_F$,  the number of fermions, do not mix with odd $N_F$ states. Now, the Hamiltonian can be diagonalized in either even or odd fermion number sector separately. We confine ourselves to even sector, as the ground state belongs to even fermion sector for $N$ even and periodic boundary conditions. Now the
bilinear terms in fermion operators can be uncoupled to a degree by going over the momentum space variables, the fermion operator is written as,
\begin{equation}
c_l= {1\over N} \sum_k e^{iql}c_k,
\end{equation}
where the sum is over all momentum values in the first Brillouin zone . For the case of $N, N_F$ even, and periodic boundary condition,  $N$ allowed values for $k$ are given by,
\begin{equation}
k= \pm {\pi\over N} (1,3,5,..N-1),
\end{equation}
the allowed range in the thermodynamic limit is $-\pi<k<\pi$. The Hamiltonian now has the form in terms of the momentum-space operators,
\begin{equation}
H= \sum_{-\pi<k<\pi} \varepsilon_k \chi_k (\psi_k+\psi_{k+\pi})+ h \chi_k \psi_k,
\end{equation}
where $2 \varepsilon_k=J_x e^{-ik}+J_y e^{ik}$, and $\chi_k=c_{-k}^\dag - c_k, \psi_k=c_k^\dag+ c_{-k}$. We can see that the operators with momentum values $k-\pi,-k,k, \pi-k$ are still coupled in this form. 

Let us  define a set of four operators labeled by  a positive $q<\pi/2$, as linear combinations of the fermion creation and annihilation operators,  
\begin{equation}
F_{\pm}={c_{q-\pi}\pm c_{-q}^\dag\over \sqrt{2}}, ~
G_{\pm}={c_q\pm c_{\pi-q}^\dag\over \sqrt{2}}. 
\end{equation} 
Using these combinations, the Hamiltonian can be written as,
\begin{equation}
H=2 \sum_{0<q<\pi/2} H_q,
\end{equation}
 where $H_q$ a is four-dimensional operator for a given mode. In the matrix form, using $\varepsilon\equiv \varepsilon_{1q}+i \varepsilon_{2q}$, we have
\begin{equation}
H_q=  (F_+^\dag~ G_+^\dag~ F_-^\dag~ G_-^\dag)  \left [ \begin{array}{cccc}-2 \varepsilon_{1q}&-2i  \varepsilon_{2q} & h&0\\ 
2 i \varepsilon_{2q}& 2  \varepsilon_{1q} & 0 &h\\
h&0&0&0\\
0&h&0&0  
\end{array} \right ] \left (\begin{array}{c} F_+\\G_+\\F_-\\G_- \end{array} \right ).
\end{equation}
The four eigenvalues of the above matrix are for each $q$,
\begin{equation}
\lambda_{n_1 n_2}= n_1 |\varepsilon_q|+ n_2\sqrt{|\varepsilon_q|^2+h^2},
\end{equation}
and the corresponding eigenvectors  $V_{n_1 n_2}$ are easily found, and the eigenmode operators $\zeta_{n_1 n_2}$ are constructed as linear combinations of
the above operators. In terms of the diagonal modes, the Hamiltonian takes the diagonal form,
\begin{equation}
H_q=\sum_{n_1=\pm,n_2=\pm} \lambda_{n_1 n_2}~ \zeta_{n_1 n_2}^\dag \zeta_{n_1 n_2}.
\end{equation}
The lowest energy state is given by occupying all negative energy states for each mode;  we have $\lambda_{--},\lambda_{+-}<0$ and the other two are positive. Thus, the ground state
energy is given by
\begin{equation}
E_g= -4 \sum_{0<q<{\pi\over2}} \sqrt{ |\varepsilon_q|^2+h^2}.
\end{equation}
The ground state is  non-degenerate,  and it is given by a direct product over the lowest-energy states constructed from the vacuum state for each mode, we have
\begin{equation}\label{gs}
|g\rangle= \prod_{0<q<{\pi\over2}} \zeta_{--}^\dag \zeta_{+-}^{\dag} |\rm vac\rangle,
\end{equation}
The normal-mode operators are given by, using the eigenvectors $V_{n_1 n_2}$,
\begin{eqnarray}
\zeta_{--}^\dag=&{\cal C} F_+^\dag+i{\cal S}E_+^\dag+h_1{\cal C}F_-^\dag+ih_1{\cal S}E_-^\dag , \nonumber\\
\zeta_{+-}^\dag=&{\cal S}F_+^\dag-i{\cal C}E_+^\dag+h_2{\cal S}F_-^\dag+ih_2{\cal C}E_-^\dag, 
\end{eqnarray}
where we have parametrized $\varepsilon_{1q}=|\varepsilon_q|\cos{\theta},\varepsilon_{2q}=|\varepsilon_q|\sin{\theta}$, and ${\cal C} = \cos{\theta/2} , ~ {\cal S} = \sin{\theta/2}$,
and $h_1=h/\lambda_{--},h_2=h/\lambda_{+-}$.\\

\subsection*{Ground state degeneracy for $h=0$}
\noindent We can see that for the case of $h=0$,  a macroscopic degeneracy arises from the eigenvalues $\lambda_{n_1 n_2}$. For each mode, the eigenvalues are
$-2|\varepsilon_q|, 0, 0, 2|\varepsilon_q|$, the zeroes correspond to $\lambda_{+-},\lambda_{-+}.$ This means, the lowest energy state is four-fold degenerate;
each of the four states $|1\rangle =\zeta_{--}^\dag|vac\rangle, |2\rangle =\zeta_{--}^\dag\zeta_{-+}^\dag|vac\rangle, |3\rangle =\zeta_{--}^\dag\zeta_{+-}^\dag|vac\rangle,|4\rangle =\zeta_{--}^\dag\zeta_{+-}^\dag\zeta_{-+}^\dag |vac\rangle$ has energy $-2|\varepsilon_q|$. This even-odd symmetry in the lowest energy state exists for every mode. This translates to $4^{N/4}$ degenerate states for the ground state.  Half  of these
states have even number of fermions, and the other half of states belong to odd sector. The physical
states should have even number of fermions, as we are working with even sector. To choose a physical
state,  we  first choose any of the four states for $N/4-1$ modes, 
which gives us either an even or an odd state. If the state is even,  then we choose either $|2\rangle$ or $|3\rangle$ for the last mode to make it a physical state,  else, we choose one of
the other two odd states to make the total state  even in fermion number. Thus the ground state degeneracy $D_g$ is given by
\begin{equation}
D_g= 2^{{N\over 2}-1}.
\end{equation}
 
\section{Operator Schmidt decomposition and Block entanglement entropy}

We are now interested in the entanglement distribution in the ground state of the spin chain, as a function of the coupling strengths and the magnetic field.
The entanglement block entropy is expected to follow an area growth law in general, and exhibit a faster growth  behavior as a function of the block size $L$ near a critical point. For one-dimensional
system, this implies, a size-independent behavior away from critical regions (where all correlations are short-ranged) and a logarithmic behavior near a critical point\cite{plenio}. 
The ground state written
using the normal-mode fermion operators, as given in Eq.12, has a Slater-determinental wave function. With these wave functions, the reduced density matrix for a block of spins, and
the entanglement  can be calculated using the correlation function matrix\cite{latore}. We show a method here that can  calculate the Schmidt eigenvalues of the state directly, 
using  a operator Schmidt decomposition of the exponentiated operator that generates the ground state. 

Let us briefly recall the Schmidt decomposition of a many-qubit pure state.
Let us bipartition the spin chain, part $A$ contains the  $L$ contiguous spins  starting with the first spin, the rest of  the spins, $L^\prime=N-L$ contiguous spins form
the second  part $B$, let $L<N/2$. Any pure state of the multi-spin state is written as,
\begin{equation}
|\psi\rangle = \sum_{i=1}^{2^L} \sum_{j=1}^{2^{L^\prime}} \gamma_{i,j}~|i\rangle_A |j\rangle_B,
\end{equation}
where $\gamma_{i,j}$ is the wave function amplitude for the product basis state of the composite system. We can view this coefficient as an element of a matrix $\hat \gamma$ which
has dimension $2^L \times 2^{L^\prime}$.  
 The above state can be written as sum over at most $2^L$ terms, in the Schmidt-decomposed form, as
 
\begin{equation}
|\psi\rangle = \sum_{k=1}^{2^L} \sqrt{\Lambda_k}~|k\rangle_A |k\rangle_B,
\end{equation}
where the Schmidt numbers $\Lambda_k$ are the eigenvalues of $\hat \gamma \hat \gamma^T$, where $T$ denotes the transpose of the matrix. The reduced density matrix
$\rho_A= tr_B |\psi\rangle \langle \psi|$, calculated by a partial trace over $B$ states, has the above basis states as the eigenstates, and the Schmidt numbers as the
eigenvalues. Thus, the entropy of the block $A$,  which is a measure of the entanglement between the two parts,  is given by
\begin{equation}
E_{AB} =-Tr \rho_A \log \rho_A = - \sum_k \Lambda_k \log \Lambda_k
\end{equation}

\begin{figure}
\vglue -12cm
\hglue -3.5cm 
\includegraphics[width=22cm,height=30cm]{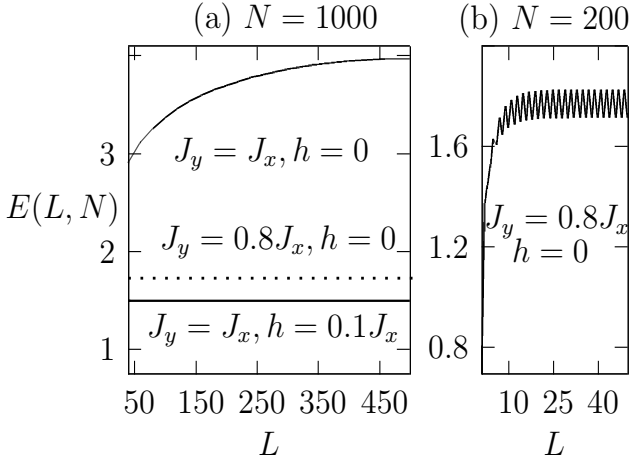}
\vglue -11cm
\caption{(a) The block entropy $E(L,N)$ of a block of $L$ contiguous spins is plotted as a function of the size for different set of values for the interaction strengths and the transverse magnetic field. (b) The entropy 
for $J_y\ne J_x, h=0$,  for $N=200$, exhibits an even-odd oscillatory variation. }
\end{figure}
Now, the calculation of the eigenvalues $\Lambda_k$ for the ground state (given in Eq.\ref{gs}) can still be a daunting task, as this entails construction of the reduced density
matrix which is $2^L$ dimensional. 
We have the ground state for all values of interaction strengths written in terms of noninteracting fermion modes that are complicated linear combinations of the momentum space fermion variables defined in Eq.3. We will rewrite the state in a more transparent form below, so that the ground state can be brought to a Schmidt-decomposed form. 
After unscrambling the operators given in Eq.6, we can express the ground state as,   modulo a normalization constant, 
\begin{equation}
|g\rangle = \prod_{0<q<{\pi\over 2}} (1+ {1-h_1\over 1+h_1} c_{q-\pi}^\dag d_q^\dag) (1+ {1-h_2\over 1+h_2} c_{q}^\dag    e_q^\dag)|0\rangle,
\end{equation}
where $d_q=\cos\theta c_{-q}+i \sin\theta c_{\pi-q},~ e_q=\cos\theta c_{\pi-q}+i \sin\theta c_{-q}$. Finally, the operator product can be exponentiated, and after a further manipulation
we can write the unnormalized ground state as
\begin{equation}
|g\rangle =e^{\hat Z}|0\rangle,
\end{equation}
where the operator in the exponent is given by
\begin{eqnarray}
\hat Z= \sum_{0<q<{\pi\over 2}} {\varepsilon_{1q} \over h +\sqrt{|\varepsilon_q|^2 + h^2}} (c_{q-\pi}^\dag c_{-q}^\dag - c_q^\dag c_{\pi-q}^\dag) \nonumber\\ +   {i\varepsilon_{2q} \over h +\sqrt{|\varepsilon_q|^2 + h^2}} (c_{q-\pi}^\dag c_{\pi-q}^\dag - cq^\dag c_{\pi-q}^\dag).
\end{eqnarray}
We manipulate from this  form of the state in the next section , to bring it to a Schmidt decomposed form to calculate the block entanglement entropy. We employ a method of Schmidt decomposing the operator $\hat Z$ itself,  which is tantamount to calculating the eigenvalues of a $L-$dimensional matrix. Let us express  the operator in real space basis as,
\begin{equation}
\hat Z = \sum_{l=1}^N \sum _{m=1}^N \gamma_{l,m}~ c_l^\dag c_m^\dag,
\end{equation}
where the coefficient $\gamma_{l,m}$ is the Fourier transform of the coefficients in Eq.17. In terms of the coefficients $\beta_1,\beta_2$ given by,
\begin{equation}
\beta_n(x)= {1\over N} \sum_{q<\pi/2} {\varepsilon_{nq}\over h+\sqrt{|\varepsilon_q|^2 +h^2}} e^{ikx},
\end{equation}
we have $\gamma_{l,m} = \beta_1(l-m) [ (-1)^l - (-1)^m ]+ i  \beta_2(l-m) [ (-1)^{l+m} -1]$. Now, the exponentiated operator is written in real space. We can now partition the operator into three parts as,
\begin{equation}
\hat Z= \hat Z_A + \hat Z_B + \hat Z_{AB},
\end{equation}
where the the first (second) term involves  operators from part $A (B)$ only. The last entangling term involves one fermion operator each from the two parts. All these three terms commute with
each other. It is easy to see that the exponentiated operator $e^{\hat Z_A+\hat Z_B}$ generates a product of local transformations on the two parts;  it will not
generate any entanglement. Thus,  we may drop the operators that act only on part A (B) only.
That is, the ground state $|g\rangle$, given in Eq.19, and the state given by
\begin{equation}\label{expo}
|\tilde g\rangle = e^{\hat Z_{AB}}|0\rangle,
\end{equation}
will have the same entanglement properties, as they are related by a product of local  transformations.  Let us define more-suggestive fermion operators,
$A_l^\dag= c_l^\dag,$ for $l=1,L$, and $B_l^\dag=c_{l+L}^\dag$, for $l=1,N-L$. We can write the operator appearing above as,
\begin{equation}
\hat Z_{AB} =  \sum_{l=1}^L \sum_{m=1}^{N-L} \Gamma_{lm} A_l^\dag B_m^\dag
\end{equation}
where the coefficient is given by,
\begin{eqnarray}
\Gamma_{lm}=2  [ (-1)^l - (-1)^{m+L} ]~ {\rm Re}~ \beta_1(l-m-L)~\nonumber\\ -~ 2   [ (-1)^{l+m+L} -1]~ {\rm Im}~ \beta_2(l-m-L).
\end{eqnarray}
Now, we can employ an operator Schmidt decomposition, in analogy with the state decomposition of writing the state as in Eq.16 from Eq.15, we have
\begin{equation}
\hat Z_{AB} = \sum_{n=1}^L \sqrt{\eta_n} A_n^\dag B_n^\dag,
\end{equation}
where the operator Schmidt numbers $\eta_n$ are eigenvalues of the matrix $\hat\Gamma \hat\Gamma^T$, and the new operators $A_n$ and $B_n$ are linear combinations of
the original site basis operators. Now, the normalized equivalent ground state is a direct product, given by
\begin{equation}
|\tilde g\rangle = \prod_{n=1}^L (x_n+ y_n A_n^\dag B_n^\dag) |0\rangle,
\end{equation}
where $x_n= 1/\sqrt{1+\eta_n}$ and $y_n= \sqrt{\eta_n}/\sqrt{1+\eta_n}$. The above has the Schmidt decomposed form (as in  Eq.16), with $2^L$ terms in the expansion, the amplitude
for each term is a product of $L$ factors.  The  eigenvalues of $\rho_A$ are labeled by numbers $\xi_n$ (equal to  0 or 1 corresponding to either $x_n$ or $y_n$ appearing in each term of the expansion),  we have
\begin{equation}
\Lambda(\xi_1,..\xi_L)= \prod_{n=1}^L |x_n|^{2\xi_n} |y_n|^{2(1-\xi_n)}.
\end{equation}
The entropy of $\rho_A$, viz. the block entanglement $E(L,N)$,  is straightforward to evaluate, and we get,
\begin{equation}
E(L,N)=  \sum_{n=1}^L H({1\over 1+\eta_n}),
\end{equation}
where the Shanon binary entropy function is given by $H(x)=-x \log x - (1-x) log (1-x)$.

This completes the method of expressing the entanglement entropy in terms of operator Schmidt numbers, which  gives an efficient method for numerical computation of the block entanglement. 
It can be easily employed for other spin chain models that are solvable by Jordan-Wigner fermionization, where the ground state is amenable to exponentiation, as in
Eq.\ref{expo}.
However, the method is not applicable for the Bethe-ansatz state of the Heisenberg-XY spin chain, as it is not clear how to exponentiate it.  We expect the method to work for Kitaev honeycomb lattice, and other higher-dimensional systems, as long as the ground state is expressed as a direct product over modes, and exponentiation can be done, as in Eq.\ref{expo}. Some well-known states of the interacting electron systems, like the BCS state of the
superconductors, have a similar structure. 
This method reduces the numerical effort substantially in the study of the eigenvalues of the reduced density matrix for large lattices. It will be interesting to employ this method to study the entanglement spectrum, and the phase diagram of the electron/spin models from the entanglement perspective.
\begin{figure}
\vglue -12cm
\hglue -3.5cm 
\includegraphics[width=21cm,height=30cm]{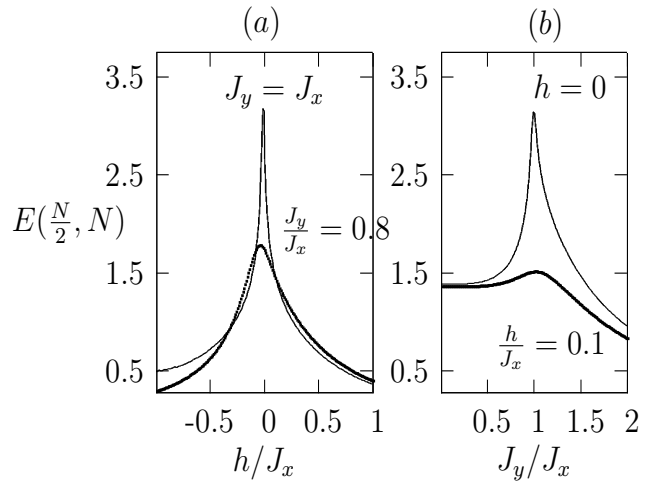}
\vglue -11.5cm
 \vglue -0.75 cm
 \caption{The block entropy $E({N\over 2},N)$ of a block of $L={N\over 2}$ contiguous spins, for a closed chain with $N=200$ spins, as a function of  (a) the magnetic field (b) the interaction strength.}
\end{figure}

\section{Discussion of Results}

We now turn to the block entanglement entropy for the ground state of our model as a function of the interaction strengths, the transverse field and the block size.  
We make a bipartition of the lattice of $N$ sites in to two blocks of size $L$ and $N-L$.
Though the parent state is pure of the composite system, the block reduced density matrix may not represent a pure state, thus it carries an entropy.  The amount of
quantum correlations present in the state is quantified by the bipartite von Neumann entropy.
Since the composite state is a pure state, it suffices to study the block entropy $E(L,N)$ as a function of the size $L$ for $L\le N/2$, for all values of the interaction strengths and the transverse magnetic field. The scaling of the entropy with the block size $L$ is related to the quantum correlations of the system, viz. the entropy is
expected to follow the area law\cite{srednicki,vidal} when the correlations are short ranged or the system is away from criticality. 
For the one-dimensional system, this implies that the entropy is independent of the size (since the block has a surface with just two sites irrespective of the size), as we will 
see below this is the case for all values of the interaction strengths when a nonzero field is present. The deviation from the area law is expected, with a logarithmically
increasing entropy as a function of the block size, in the vicinity of a critical point. Since the correlations tend to be long ranged near a quantum critical point,  they should reflect in behavior of the entropy as well. It has been shown using conformal field theories, which are related to the continuum limit of the gapless spin systems near
a critical point, the coefficient of the logarithmic correction of the entropy is  related to the central charge of the system. We will see below that the spin chain we are
investigating shows a critical behavior for $J_x=J_y, h=0$, and for all other values it is gapped.

We now turn to the numerical computation of the block entropy using the method we developed in the previous section.
The numerical task is to get the operator Schmidt numbers of Eq.27, from the diagonalization of a $L$-dimensional matrix  $\Gamma\Gamma^T$  (see Eq.25). This
means we can consider spin chains with a large number of sites.
To this end, we have numerically computed the operator Schmidt numbers from Eq.25,  in the various parameter regimes,  for a large spin chain with $N=1000$ spins.  Fig.1a shows the block entanglement entropy as a function
of the block size $L$ for different values of the interaction strengths and the magnetic field.  We have shown here the block entropy for the case of even $L$ only. This
is because there is even-odd oscillatory behavior of the entropy with the size.

As we can see from Fig.1a, 
at the macroscopic degeneracy point,  for $h=0, J_y=J_x$, the entanglement entropy exhibits a logarithmic behavior for $L<N/2$, which is an expected behavior for
a system in the vicinity of a critical point. The entropy is  given by
\begin{equation}
E(L,N)\approx {c\over 3 } \log_2{L},~~~ {\rm for}~ J_y=J_x, ~h=0,
\end{equation}
the coefficient, known as the central charge, seems to fit $c={3\over 2} \log{2}$. This value of the central charge 
is different from other spin models\cite{plenio,vidal}, for example $c=1$ for the Heisenberg spin chain. We can take that the spin correlations are long-ranged
at this degeneracy critical point, in analogy with the conformal field theory result for other spin models with known critical-point behavior of the correlation functions.

As we can see from Fig.1a, for $h\ne 0$ the block entropy is independent of the block size,  thus
obeying the area law, implying the system is away from the critical region. This is expected as the nonzero transverse field lifts the macroscopic degeneracy, shown in  Eq.14, we expect the system to be away from criticality, presumably with short-ranged correlations. However, we see from Fig.1a that the entropy obeys the area law even for
$h=0$ but the interaction strengths are not equal, $J_x\ne J_y$. Here, the ground state still has a macroscopic degeneracy, just as for equal interaction strengths. 
In this case we seem to have conflicting indications, the entanglement entropy obeying the area law indicates the system is away from a critical point with
short-ranged correlations, the gapless excitation spectrum accompanying a macroscopic degeneracy indicates the system is in the vicinity of a critical point accompanied
by long-ranged correlations.  Similar conflicting features have been seen in the two-dimensional Kitaev model by Baskaran, Sen and Shankar\cite{general},
where a parameter regime exists in which the ground state has gapless excitations along with short-range correlation functions.
The present situation may be similar, and we conjecture that off-diagonal correlation functions may be short ranged in this model. The diagonal correlations may be
long ranged due to a spin polarization that can result due to the unequal interaction strengths. 
The block entropy
will depend on the diagonal and off-diagonal correlations functions and their range,  analogously of the other measures of the entanglement like concurrence\cite{subrah}.
The block entanglement, unlike the concurrence measure, will depend on various multi-spin correlation functions, making the situation more complicated. Since our
method deals directly with the eigenvalues of the reduced density matrix only, a detailed study of multi-spin correlations cannot be addressed here.

In the regime with unequal interaction strengths and zero magnetic field, the entanglement entropy exhibits an even-odd non-monotonic behavior as a function of
the size.
The oscillatory behavior of the block entanglement is highlighted in Fig.1b for a smaller chain with $N=200$ spins.
This  even-odd effect is persistent for very large sizes also, as the block entanglement itself becomes size independent in this
regime. This non-monotonic behavior  may be related to the fact that we have an inhomogeneous chain, and for odd $L$, the block will have unequal number
of $J_x$ and $J_y$ bonds. Similar behavior has been seen in spin-one chains with bi-quadratic interactions, in spin chains with boundaries, in SU(n) Hubbard chains with further neighbor hopping\cite{oscillate}.   


We can also study the entropy as a function of the interaction strengths and the magnetic field separately, to track the critical-point behavior and its signature. We fix
the block size to be $L=N/2$, so the entropy takes its maximum value.
In Fig.2a, we have plotted the block entropy of the largest block,  for $N=200$, $L=N/2$, as a function of $h/J_x$ for the case of $J_y/J_x=1$ and $J_y/J_x=0.8$.
As a function of the magnetic field, the block entropy displays a peak structure at $h=0$, for $J_y=J_x$, tracking the quantum critical behavior at $h=0$. But for the case of unequal strengths,
$J_y/J_x=0.8$, the peak structure is quite diminished, apart from shifting from $h=0$, to a slightly negative value. Away from the peak, the entropy itself becomes small for
large magnetic field. The spins are expected to be polarized predominantly along the field direction, thus decreasing the entropy. 
We have shown in Fig.2b the behavior of the block entropy as a function of $J_y/J_x$, for both $h=0$ and $h\ne0$. Here also, the block entropy tracks the critical behavior with a sharp peak near $J_y=J_x$ for
the case of $h=0$. For large $J_y$, for a fixed $J_x$ and a small field, the spins are polarized along $y$ direction, thus decreasing the entropy. 
In both the cases of a nonzero magnetic field and/or unequal interaction strengths , the system is away from critical point, as we discussed earlier in the context of
the size dependence of the block entanglement.  The peak structure in this case cannot really be associated to the tracking  of a critical point, as a true critical
behavior obtains only for $h=0, J_x=J_y$.
The diminished peak may signify a smear or rounding off in the critical behavior, and thus can be useful in locating a critical point in numerical computations by tuning
the parameter regime. 

In conclusion, 
we have shown that the spin chain with Kitaev-type interactions is exactly solved through Jordan-Wigner fermionization. We have employed a method that reduces the
task of finding $2^L$ eigenvalues of the reduced density matrix of a block of $L$ spins,  to finding the eigenvalues of a $L-$dimensional matrix. Using this method, we have 
numerically computed block entropies for the spin chains with a large number of spins. The block entropy obeys the area law for all interaction strengths for a nonzero
transverse magnetic field, implying the system is gapped and away from a critical point. The block entropy shows a logarithmic increase with the block size only at
the degeneracy point $h=0, J_x=J_y$, though there is a macroscopic degeneracy for $J_x\ne J_y$.

It is a pleasure to acknowledge Prof. R. Shankar for an insightful discussion about the Kitaev spin chains.

 \end{document}